\documentstyle[preprint,tighten,prb,aps]{revtex} 
\begin{document}
\draft
\preprint{Nota Cientifica PUC-Rio 23/91}

\title{Ising Spin Glass in a Transverse Magnetic Field}

\author{Beatriz Boechat, Raimundo R. dos Santos\cite{email}}
\address{Departamento de F\'\i sica,
         Pontif\'\i cia Universidade Cat\'olica do Rio de Janeiro,\\
         Cx Postal 38071, 22452-970 Rio de Janeiro, Brazil}
\author{M. A. Continentino}
\address{Instituto de F\'{\i}sica, Universidade Federal
         Fluminense,\\
         Outeiro de S\~ao Jo\~ao Batista, s/n, 24210 Niter\'oi,
         Brazil}
\date{24 November 1993}
\maketitle
\begin{abstract}
We study the three-dimensional quantum Ising spin glass in
a transverse magnetic field
following the evolution of the bond probability distribution
under Renormalisation Group transformations.
The phase diagram
(critical temperature $T_c$ {\em vs} transverse field $\Gamma$)
we obtain shows a finite  slope near $T=0$, in contrast
with the infinite slope for the pure case.  Our results
compare very well with the experimental
data recently obtained for the dipolar Ising spin glass
LiHo$_{0.167}$Y$_{0.833}$F$_4$, in a transverse field.
This indicates that this system is more apropriately described by
a model with short
range interactions than by an equivalent Sherrington-Kirkpatrick
model in a transverse field.

\end{abstract}
\bigskip

The role of quantum fluctuations in spin glasses has been a long
standing theoretical problem\cite{Klemm,Bray,dosSantos85}.
The so-called proton glasses\cite{Akheizer83,Courtens84,Pirc87}
--- a
random mixture of ferroelectric and antiferroelectric materials
such as Rb$_{1-x}$(NH$_4$)$_x$H$_2$PO$_4$ --- provide an
experimental realization for quantum spin glasses.
Within a pseudospin description of such hydrogen-bonded
systems, the proton position in the two potential minima is
represented by Ising states, $\sigma^z=\pm1$, and the tunneling
between the minima by a transverse field term, $\Gamma\sigma^x$,
where $\Gamma$ is the tunneling frequency\cite{deGennes63}.
The theoretical study of Ising spin glasses in a transverse field
(TISG) has then attracted renewed interest, especially in relation
to the analogue of the Sherrington-Kirkpatrick model\cite{SK} in a
transverse field (TSK); see, {\em e.g.,} Ref.\ \onlinecite{Yadin}
 and
references therein. More recently, the magnetic susceptibility
of the (long-ranged dipolar) Ising spin glass
LiHo$_{0.167}$Y$_{0.833}$F$_4$ has been measured in the presence
of a transverse field $H_t$, from which a phase diagram
$T_c(H_t)$ was determined\cite{Aeppli1,Aeppli2}.
Therefore, it is of interest to discuss the main 
differences between the phase diagrams of the
transverse Ising model in both pure and spin-glass cases.
In view of the long range nature of the interactions in
the dipolar glass, we are also particularly interested in establishing
whether this system can be suitably described by a short-range
model or one has to resort to the TSK model.
Here we address these questions using
real-space scaling methods.

The TISG model is described by the Hamiltonian:
\begin{equation}
-\beta{\cal H} = \sum_{i<j}^NJ_{ij}{\bf\sigma}^z_i{\bf\sigma}^z_j +
\Gamma\sum_i^N{\bf\sigma}^x_i
\label{H}
\end{equation}
where the $\sigma_i^\mu\ ,\mu=x,z$ are Pauli spin matrices, $\Gamma$
is the
transverse field, $i$ and $j$ are nearest-neighbor sites on a simple
cubic lattice,
and the $J_{ij}$ are uncorrelated exchange couplings
chosen at random from an even  distribution.
For zero transverse field the model reduces to the classical Ising
spin glass, and quantum effects are brought in by increasing the field.
At finite temperatures, the effect of the transverse field is to depress
the spin glass transition temperature, whereas at zero temperature quantum
fluctuations are the only mechanism driving the system to a phase
transition at a critical value of the transverse field.
It is interesting to note that the lower critical dimension ($d_\ell$)
for zero temperature transitions in the Transverse Ising spin-glass model
is \cite{dosSantos85}~$d_\ell=1$, unlike the classical Ising case
\cite{BY}, $d_\ell=3$.

In the context of real-space renormalisation group, the simple cubic
lattice may be approximated by hierarchical Migdal-Kadanoff
cells\cite{BvL}; see Fig.\ 1.
The terminal sites are connected by $b^{d-1}$ bonds `in parallel',
each of which consists of $b$ bonds `in series'; $b$ is the scaling
factor ($b=2$ in Fig.\ 1) and $d$ is the space dimensionality of the
lattice ($d=3$ in this case).
For the transverse Ising model (TIM), the non-commutation aspects are
present at the cluster level, in the sense that individual spins
are not in a definite state.
This can be dealt with by referring the density matrix to
the basis  $|m_1 m_2 \ldots m_N\rangle$, where
$\sigma_i^z|m_i\rangle=m_i|m_i\rangle$, and defining the
renormalisation group
transformation (RGT) by the mapping of diagonal
elements only. This
approach has been successfully used
in a detailed study of the pure and bond-diluted TIM in two
dimensions\cite{dosSantos}.
In the present work bond
disorder is included within a statistical renormalization group
(SRG) treatment: one follows the effect of a RGT
on the probability distributions of the relevant
parameters, instead of forcing them into a particular
form\cite{Robin}. Several aspects of the Ising spin glass\cite{Southern}
and of the random field Ising
model\cite{bia} have been elucidated by treating disorder this way.

Thus, for a given bond configuration ($\{J_{ij}\}$), an RGT for the $d=3$
system (Fig.~1) is defined by
\begin{equation}
\langle m_1m_6\vert\rho^{\prime}({\bf K^{\prime}})\vert m_1m_6\rangle =
\langle m_1m_6\vert{\tilde\rho}({\bf K})\vert m_1m_6\rangle
\label{transform}
\end{equation}
where ${\bf K^{\prime}} = (J^{\prime},
\Gamma^{\prime}, C^{\prime})$ are
the renormalized quantities in the two-site cell, ${\bf K} =
(\lbrace J_{ij}\rbrace ,\Gamma)$ refers to the original
cluster,
and
\begin{equation}
\langle m_1m_6\vert{\tilde\rho}({\bf K})\vert m_1m_6\rangle =
\sum_{m_2m_3m_4m_5}\langle m_1m_2m_3m_4m_5m_6\vert\rho({\bf K})
\vert m_1m_2m_3m_4m_5m_6\rangle
\label{diag1}
\end{equation}
is obtained by performing the partial trace on the internal spins,
keeping those on the terminal sites fixed (see Fig.\ 1).
The third  matching condition required to solve the problem for
the three unknows $(J^{\prime},
\Gamma^{\prime}, C^{\prime})$
is obtained by preserving the thermodynamical average of the
rotation operator
\begin{equation}
\langle {\bf\sigma}_1^x{\bf\sigma}_6^x\rangle_{{\cal H^{\prime}}} =
\langle {\bf\sigma}_1^x{\bf\sigma}_2^x{\bf\sigma}_3^x{\bf\sigma}_4^x
\sigma_5^x\sigma_6^x\rangle_{\cal H'}
\label{rot}
\end{equation}

For the renormalized cell, Eqs.\ (\ref{diag1}) and (\ref{rot})
provide analytical expressions for the three primed unknown
variables~\cite{dosSantos}. Since disorder destroys
the point group symmetry of the original  cluster, the
Hamiltonian matrix is written in a $64\times 64$ representation and
Eqs.\ (\ref{diag1}) and (\ref{rot}) are calculated numerically.
At this point we should comment that a single renormalized field
appears in
the recursion relations as a result of the approximation employed
here: the fields are assumed to be uniform in both renormalized
and original cells.
This assumption can be justified, to some extent, by recalling
that the transverse magnetic field behaves as an irrelevant variable
in the pure TIM. Since the transverse field is not a symmetry
breaking operator, this should hold in the spin glass case as well.
On the other hand, we could have allowed the field at each site to
follow an RG trajectory. The irrelevance of the field would
manifest itself through a distribution evolving from
an initial delta function centered at $\Gamma=\Gamma_0$ into one
centered at $\Gamma=0$.

The initial probability
distribution for the exchange couplings is
\begin{equation} P(J_i) = {1 \over \sqrt{2 \pi \tilde J^2}} \exp
\left\lbrack -{J_i^2\over 2\tilde J^2}\right\rbrack ,
\label{PJ}
\end{equation}
where we have assigned an index $i$ to each bond in the original
cell and taken $\langle J_iJ_l \rangle =\tilde J^2\delta _{il}$ and
$\langle J_i\rangle =0$.
We start the iteration by choosing eight bonds distributed
according to
Eq.\ (\ref{PJ}) and one value of the transverse field
($P(\Gamma ) = \delta (\Gamma_i-\Gamma)$), to feed the recursion
relations and generate a new value of the field and of the exchange
coupling.
This procedure is repeated about 10,000 times and we obtain two
renormalized distributions, $ P^{\prime}(J^{\prime}_i)$ and
$P^{\prime}(\Gamma^{\prime})$. We make use of these
distributions to feed the recursion relations in the next step
of the renormalization process:
eight bonds and one field
are chosen according to the new distributions, and this is done
again 10,000 times.
The evolution of the distributions is then followed along the
renormalization process.
We find that the bond probability distribution remains symmetric
around $J=0$ at each iteration; that is, $\langle J_{ij}\rangle = 0$.
The attractors of the different
phases are
determined by fixed
distributions characterized by
their width $\tilde J$ and  mean value
$\langle\Gamma\rangle$, as follows:
\newpage
$$
\tilde J \rightarrow \infty, \ \ \ \langle\Gamma\rangle \rightarrow 0
\ \ \ {\rm spin~glass}
$$
$$
\tilde J \rightarrow 0, \ \ \ \langle\Gamma\rangle \rightarrow 0 \ \
\ {\rm paramagnetic}.
$$
so that the critical curve is determined as the boundary between
these two different behaviours. Note that in every case the average
value of the transverse field distribution iterates to zero.

In Fig.\ 2 we present our results for the phase diagram, $T_c\ vs\ \Gamma$,
for both pure and spin glass cases. For
comparison we also display the spin glass data for the
replica-symmetry-breaking
(RSB) solution of the TSK
model\cite{Yadin}, and for
LiHo$_{0.167}$Y$_{0.833}$F$_4$\cite{Aeppli1,Aeppli2}.
In analysing the experimental data one should have
in mind\cite{Aeppli1,Aeppli2} that
the applied transverse field $H_t$ gives rise to a level splitting
$\sim H_t^2$ (at low fields) which, in the context of
Eq.\ (\ref{H}), is proportional to $\Gamma$;
thus, $\Gamma\sim H_t^2$.
For the pure TIM, the critical line both for small fields and near
$T=0$ has a square fit:
$T_c(0)-T_c(\Gamma)\sim\Gamma^2$, and
$\Gamma_c(T) -\Gamma_c(0)\sim T^2$, respectively.
Still for the pure case, the critical field for zero temperature
transitions obtained with the present RG
is $(\Gamma /J)_c= 3.40$, which should be compared with the
series result\cite{Yanase76}, $(\Gamma /J)_c= 5.14$; similarly,
our result $T_c(0)= 3.83J$ should be compared with
$T_c= 1.13 J$, obtained from series expansions for
the three-dimensional Ising model\cite{Domb}. As usual, the critical
parameters obtained within a simple Migdal-Kadanoff approximation
are quite inaccurate,
but one is generally able to describe the qualitative aspects of
phase diagrams\cite{BvL,Robin}.

For the transverse Ising spin glass, we obtain a curve $T_c(\Gamma)$
with a finite slope near $T=0$, unlike both the pure case
and the replica symmetry-breaking-solution to the TSK model; see
Fig.\ 2. Overall, the experimental data are better represented by the present
approach than by the infinite-range mean-field model\cite{Yadin}.
This can be explained by the fact that in the actual crystal,
the dipolar interactions fall off with the distance, being
effectively reduced to zero for distances greater than a
few lattice spacings. In contrast, the interactions between
any pair of spins  in
the mean-field model have the same intensity, irrespective of
the distance between them.
The calculated critical parameters in this case are:
$\Gamma_c=1.58J$ at $T=0$, and $T_c=0.884J$ at zero transverse
field. In agreement with the experimental results\cite{Aeppli1,Aeppli2}
we find that temperature is more effective in destroying the
spin-glass order than quantum fluctuations

We point out that the renormalisation group trajectory along the
critical line flows away from the zero temperature fixed point,
towards the one controlling the classical finite temperature
spin-glass transition. This has two main implications:

(1) The shape of the critical line close to $\Gamma = 0$ is
analytic\cite{Suzi}. In the present case our results are consistent
with $T_c(0)-T_c(\Gamma)\sim\Gamma^2$.

(2) The exponents controlling the transition for finite $\Gamma$ are
the same as those of the classical spin glass transition, except at
zero temperature where quantum effects become dominant.

We can develop a scaling theory for the spin glass transition close
to the unstable zero temperature fixed point at $(\Gamma /J)\simeq
(\Gamma /J)_c$. Introducing an exponent $z$ through the scaling
relation\cite{Mucio}
\begin{equation}
\label{scaling}
\tilde J^{\prime} = b^{-z}\tilde J,
\end{equation}
we obtain the scaling form for the free energy density close to
$(\Gamma /J)_c$ as\cite{Mucio}
\begin{equation}
f = \left\vert g\right\vert^{2-\alpha}{\cal F}\left\lbrack
{(T/J)\over\left\vert g\right\vert^{\nu z}}\right\rbrack
\end{equation}
where $g=\left\vert (\Gamma /J) - (\Gamma /J)_c\right\vert$,
 $\nu$ and $z$ are
the correlation length and the dynamic exponents, respectively,
$\alpha$ is a critical exponent which describes the singularity of
the ground state energy; all these exponents are associated with the
zero temperature fixed point and are related through the modified
hyperscaling
relation $2-\alpha = \nu (d+z)$. Close to $(\Gamma /J)\simeq
(\Gamma /J)_c$ the critical temperature
vanishes as
\begin{equation}
T_c\propto  \left\vert g\right\vert^{\nu z}
\end{equation}
which allows us to obtain the product $\nu z$. Our RG results for the
behaviour of the phase boundary near $T=0$ yield $\nu z \simeq 1.23$.
An independent calculation of the exponent $z$ through a finite size
scaling analysis\cite{DSRR} for the gap at the critical point yields
$z = 1.40$.
Using these results we obtain $\nu = 0.87$ consistent with the exact
constraint $\nu \geq 2/d$ for disordered systems\cite{Chayes}.
The finite slope of the phase boundary close to $T=0$ is a consequence
of the fact that $\nu z > 1$ for the disordered case differently from
the pure three-dimensional case where $\nu z = 1/2 < 1$.

In conclusion, we have examined the phase diagram
of the transverse Ising
spin glass model.
We have compared our data with those
obtained experimentally for the
dipolar Ising spin-glass LiHo$_{0.167}$Y$_{0.833}$F$_4$
with an applied field in the transverse direction.
In spite of  the long-range character of the interactions
between spins, this system is more apropriately described
by a model with short range interactions than by the equivalent
Sherrington-Kirkpatrick model in a transverse field\cite{Yadin}.
However  the small value of the exponent $\gamma$ associated with
the non linear susceptibility was found experimentally\cite{Aeppli2}
to be
quite different from that of the classical spin glass transition.
This remains a puzzle from the point of view of our results. Further
theoretical and experimental studies are required to clear this point.
Work is in progress to investigate the possibility of the transition
at $T=0$ being first order, and to obtain a more detailed scaling analysis
of this transition.

\acknowledgments
The authors are grateful to G. Aeppli and T. Rosembaum for useful
discussions. This work started when two of us (RRdS and MAC) were
at the Institute for Theoretical Physics of the University of
California, Santa Barbara, whose hospitality is greatly appreciated.
Financial support from the Brazilian Agencies
MCT, 
CNPq, 
and CAPES 
is also gratefully acknowledged.

\begin{figure}
\bigskip
\caption{Clusters used in the RG transformation
in three dimensions. The
 terminal sites are labeled 1 and 6.}
\label{rgt} 
\end{figure}

\begin{figure}
\bigskip
\caption{Critical temperature {\em vs} transverse field for
the Ising model.
$(\Box)$ and $(\triangle)$ denote our renormalisation group (RG)
results in the pure and spin glass cases, respectively.
The replica-symmetry-breaking (RSB, Ref.\ 9) results
$(---)$ and the experimental data $(\bullet)$ for
LiHo$_{0.167}$Y$_{0.833}$F$_4$ (Refs.\ 10 and 11) are also shown.}
\label{phased} 
\end{figure}
\end{document}